\documentclass[11pt,a4paper]{article}
\usepackage{jheppub}
\usepackage{amsmath,xfrac,xcolor,amssymb,float,cancel}
\usepackage[toc,page]{appendix}

\title{Production of Primordial Black Holes in the Double $D3$-Brane - anti-$D3$-Brane Inflation Model}

\author[]{S.-H. Henry Tye}
\affiliation[]{Department of Physics,
Cornell University, Ithaca, NY 14853, USA}

\affiliation[]{Department of Physics and Jockey Club Institute for Advanced Study,
Hong Kong University of Science and Technology, Clear Water Bay, 
Hong Kong, SAR, China}

\emailAdd{tye.henry@gmail.com}

\date{8/07/26}

\abstract{ A natural extension of the successful $D3$-${\bar D}3$-brane inflation model in string theory is one with two $D3$-${\bar D}3$-brane pairs, giving rise to a novel feature in string-theory. When the first pair annihilates, $D1$-strings and $F1$-strings are produced. Following the non-commutative geometric properties in string theory, it is shown by Myers~\cite{Myers:1999ps,Johnson:2003cvf} that the presence of the $D3$-charge field strength of the remaining $D3$-${\bar D}3$-brane pair leads to the formation of dielectric $3$-branes -- neutral, finite-size bound states of $3$-branes with $D1$-strings. They behave as matter with little or no pressure. Spanning a wide range of masses, they seed primordial black holes in the early universe.

}

\begin{document}

\maketitle

\section{Introduction}\label{intro}

In the inflationary universe perspective~\cite{Guth:1980zm,Linde:1981mu}, brane inflation~\cite{Dvali:1998pa} in string theory is natural. The simplest scenario is $D3$-brane-anti-$D3$ (${\bar{D}3}$)-brane inflation in Type IIB string theory~\cite{Dvali:2001fw,Burgess:2001fx}. The workable version of the $D3$-${\bar D}3$-brane inflation model, namely, the KKLMMT model~\cite{Kachru:2003sx}, predicts the power spectral index $n_s$ (independent of any of the parameters of the model) in excellent agreement with the PLANCK Cosmic Microwave Background (CMB) anisotropy data~\cite{Planck:2016}. So it is worthwhile to study the model further. One way to test the model is to see if it can explain some challenging cosmological puzzles related to the early  universe.

Properties of black holes play a central role in the study of theoretical physics, astrophysics and cosmology. It is amazing that black holes in our universe covers a very wide range of masses. However, it is difficult to see how standard astrophysical processes alone can produce supermassive black holes, with masses up to $10^{10} M_{\odot}$ at relatively high redshifts (cf.\cite{Wikipedia}). Many believe that these black holes are primordial black holes (PBHs) formed in the early universe through a non-stellar mechanism (see reviews~\cite{Escriva:2022duf,Carr:2023tpt,Carr:2026hot}). The production of PBHs should be a natural consequence in a viable inflationary scenario, so the production (or non-production) of (super-)massive PBHs provides a way to test this $D3$-${\bar D}3$-brane inflation model, or, in our case, its simple extension.

In the original $D3$-${\bar{D}3}$-brane inflation model, the $3$-branes expand exponentially in the 3 large dimensions while the gravitational and the attractive $D3$-forces bring them together (inside the extra 6 compactified dimensions) until they collide and annihilate each other.  After the end of inflation, when the $D3$-${\bar{D}3}$-brane pair has annihilated, the energy released goes to $D1$-strings (i.e., $D1$-branes) and $F1$-strings, where some of the light modes become fundamental particles heating up the universe, while a few long strings form a cosmic string network~\cite{Jones:2002cv,Sarangi:2002yt,Copeland:2003bj}. Such a cosmic string network is consistent with present day astrophysical constraints while testable in the near future.

\begin{figure}\label{DoubleFig}
 \begin{center}
  \includegraphics[width=5.5in]{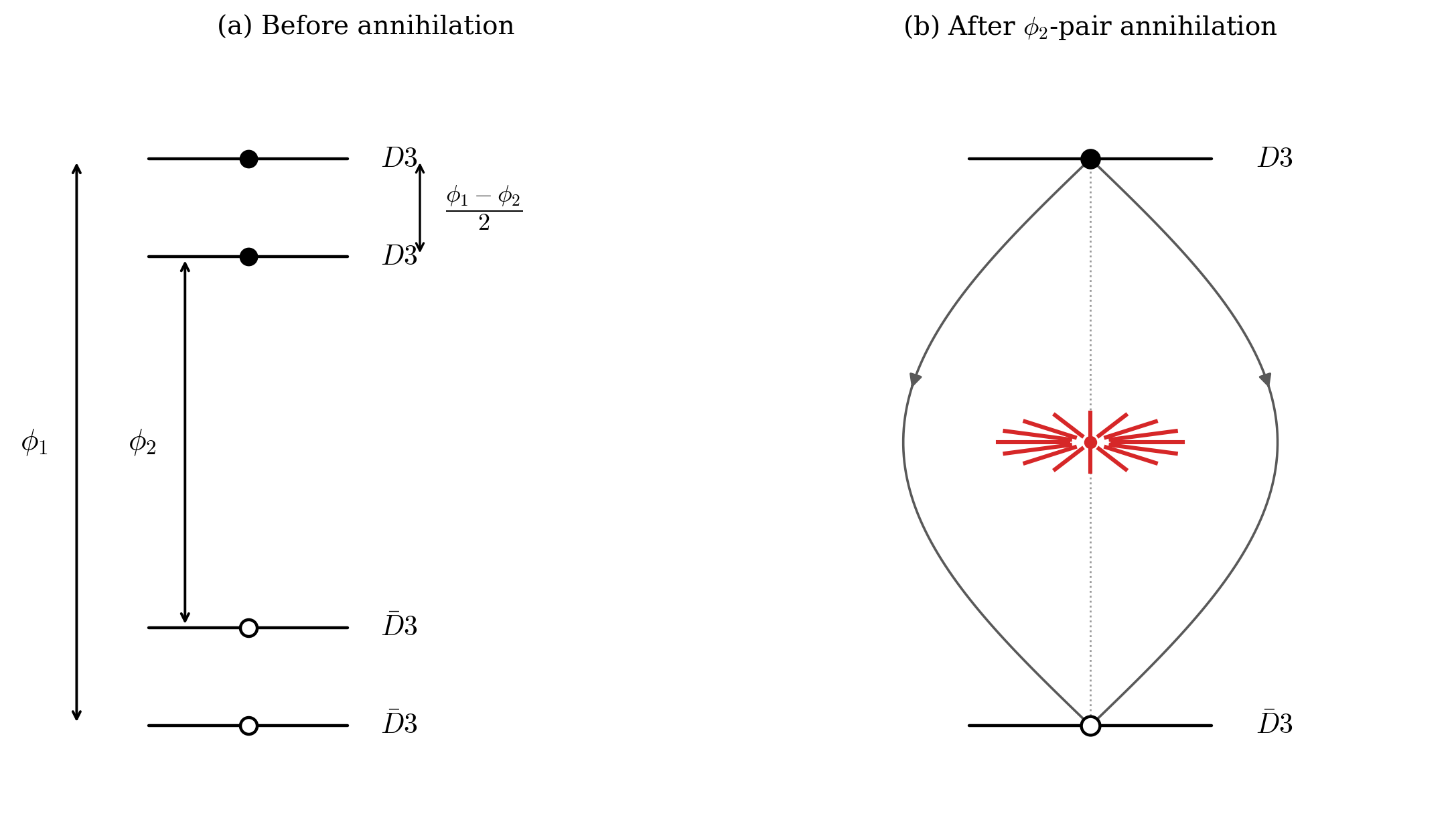}
  \caption{The double $D3$-${\bar{D}3}$-brane model in an idealized setup, where the 2 $D3$-branes (black dots) and the 2 ${\bar{D}3}$-branes (circles) are aligned along a line, with separations $\phi_1$ (outer pair) and $\phi_2$ (inner pair) as shown in panel (a). The inner $\phi_2$-pair collides and annihilates first, shown as a red flash in panel (b). $D1$-strings (and $F1$-strings) are produced in the presence of the $D3$-charge field strength (shown as flux lines) of the $\phi_1$-pair. This field strength leads to the formation of Diel-3-branes, bound states of $3$-branes with $D1$-strings. They act as matter with little or no pressure. Other arrangements of the branes are discussed in the text. 
 } 
   \end{center}
    \end{figure}

With this clear physical picture, the extension of the model to 2 pairs of $D3$-${\bar{D}3}$-branes is quite natural and relatively straightforward.  Let us outline how the model with 2 pairs of $D3$-${\bar{D}3}$-branes provides a string theory mechanism to produce PBHs in the early universe. The key difference comes when one brane pair annihilates earlier than the other pair, as is most likely the case. Here, the $D1$-strings copiously produced by the $\phi_2$-pair annihilation are in the presence of the $\phi_1$-pair, i.e., in the presence of the field strength of $D3$-charge force that attracts the remaining $D3$-brane towards the ${\bar{D}3}$-brane. In the presence of this $G_5$ field strength, as pointed out by Myers~\cite{Myers:1999ps} and with a simple generalization (cf.\cite{Johnson:2003cvf}), a collection of $D1$-strings forms a neutral dielectric $3$-brane (Diel-3-brane) with ${\cal{R}} \times S^2$ topology. Such a Diel-3-brane is a bound state of a $3$-brane with ${\cal{N}}$ $D1$-strings, where ${\cal{N}}$ can take a wide range of values. 
 
 It may be formed during the collision when a fragment of the $D3$-brane merges with the ${\bar{D}3}$-brane (or vice versa). A Diel-3-brane may take different shapes, including a 3 dimensional ball, a dumbbell and a network. They behave like matter with little or no pressure. Although we expect only a small fraction of the energy goes into the Diel-3-branes, some of them can be very massive; so we expect the Diel-3-branes formed to cover a wide range of masses. They provide the seeds for PBHs in the early universe. 

The Myers mechanism crucially depends on the presence of the 5-form field strength $G_5$. To simplify the discussion, we study the idealized setup in Fig.~\ref{DoubleFig}, which maximizes its magnitude $f =|G_5|$.
We find that the resulting scalar power spectral index $n_s$ is fully consistent with the PLANCK data~\cite{Planck:2016}. In general, the 2 pairs are not expected to line up so neatly as shown in Fig.~\ref{DoubleFig}. Moving one pair slightly away from the other pair, the $D1$-strings are produced in the presence of the dipole force with a weaker $G_5$. On the other hand, if the 2 brane pairs are far apart, then $G_5$ will be too weak for the Myers mechanism to function. It is therefore very encouraging to know that, as shown below, 
the resulting $n_s$ in this unfavorable setup is inconsistent with the PLANCK data. 

In the single $D3$-${\bar{D}3}$-brane inflation model, there are essentially two parameters, namely the $D3$-brane tension $\tau_3$ and the initial inflaton value $\phi_{ini}$ or, equivalently, the number of e-folds $N_e$ of inflation required (e.g., $N_e\simeq 50-60$). In this double $D3$-${\bar{D}3}$-brane inflation model, we have at least two more parameters, namely, the e-folds $N_2$ ($< N_e$) of inflation for the $\phi_2$-pair and the $G_5$ field strength $f$ at the spot where the $\phi_2$-pair annihilates. 
Even though the double $D3$-${\bar{D}3}$-brane inflation model is well defined,  the dynamics is non-trivial and only some of the features are explored here. In particular, we are unable to find the Diel-3-brane mass distribution function, which is very sensitive to the starting setup. Presumably, a numerical simulation with a deeper analysis will be able to address this problem. In this sense, this proposal on the production of supermassive PBHs should be treated as a conjecture.
However, although the idealized discussion here is a bit simplistic, the qualitative features of the model are robust and deserve further investigation.

The rest of the paper is organized as follows.
 Sec.~{\ref{single}} reviews the properties of the original $D3$-${\bar D}3$-brane inflation model.
Sec.~{\ref{double}} considers the general properties of the double $D3$-${\bar{D}3}$-brane inflation scenario. In the idealized case shown in Fig.~\ref{DoubleFig}, the branes are lined up along a line, and the resulting power spectral index is shown to be in excellent agreement with PLANCK data. Other arrangements of the branes and the resulting $G_5$ field strength are discussed.
Sec.~{\ref{diDp}} discusses the formation of Diel-3-branes. We first review the formation and properties of dielectric $(p+2)$-branes, and then focus on the $p=1$ case and the dual of the $p=0$ case.
Sec.~{\ref{diD3M}}  applies the dielectric $3$-brane dynamics to the double $D3$-${\bar D}3$-brane inflation model and discusses the types of Diel-3-branes formed.
Sec.~{\ref{PBHS}} presents an order-of-magnitude estimate of the fraction of energy released by the $\phi_2$-pair annihilation that goes into the formation of Diel-3-branes that seed the PBHs.
Sec.~{\ref{Remark}} gives an overall picture as a summary and contains a few remarks. Some details are relegated to 2 appendices.

\section{The $D3$-${\bar{D}3}$-brane Inflation Model}\label{single}

Type IIB string theory has 10 spacetime dimensions. Besides $F1$-strings, there are also $Dp$-branes for odd $p$.
In the brane world scenario, a $D3$-brane spans the 3 large spatial dimensions (the $xyz$ directions), while the other 6 spatial dimensions are compactified to a volume ${\cal V}$, which relates the string scale $M_S$ to Newton's constant $G_N = 1/8 \pi M_P^2$, so $M_P=2.43 \times 10^{18}$ GeV and $M_P^2 \sim M_S^8 {\cal V}$,
where $M_S \ll M_P$. The standard model branes live in a warped throat, called the $SM$-throat. Presumably  its mass scale is the electroweak scale $M_{EW}$.
The $D3$-${\bar{D}3}$-brane inflation takes place in another warped throat, called the $A$-throat, with the mass scale $M_A$, expressed in terms of the Regge slope $\alpha'$,
$M_A^2= 1/2 \pi \alpha'$. Here 
 $$M_{EW} \ll M_A \ll M_S \ll M_P$$
 so the warped $D3$-brane tension and the $D1$-string tension in the A-throat are given by, respectively,
  \begin{equation}\label{eq:tension}
  \tau_3= \frac{M_A^4}{2\pi g_s}, \quad \quad \tau_1= \frac{M_A^2}{g_s},
\end{equation}
where the string coupling $g_s$ is expected to lie in the range $1 \ge g_s >0$.

\begin{figure}
 \begin{center}
  \includegraphics[width=5in]{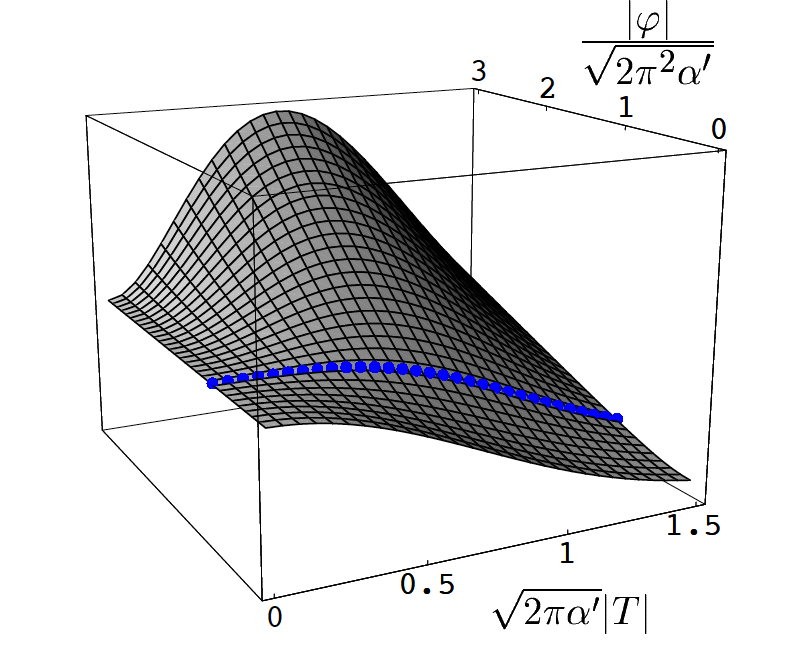}
  \caption{The $D3$-${\bar{D}3}$-brane potential where the inflaton $\phi=\sqrt{\tau_3}\varphi$ measures the separation distance between the 2 branes \cite{Jones:2002sia}. Inflation happens as $\phi$ rolls down along the $T=0$ valley. The blue line indicates the end of inflation when the complex $T$ field becomes tachyonic for $\phi \le \phi_c$. As $T$ rolls down, topological defects, i.e., $D1$-strings, are produced, in addition to the  fundamental $F1$-strings. Figure is from Ref.\cite{Jones:2002sia}}.
 \label{figureNick}
  \end{center}
    \end{figure}
    
The $D3$-${\bar D}3$-brane inflation model derived from string theory is given in Ref.\cite{Jones:2002sia}, where the full potential $U(\phi, T)$ is shown in Fig.~\ref{figureNick} (see also Ref.\cite{Tye:2023rwa}). This is a hybrid inflation model~\cite{Linde:1993cn,Garcia-Bellido:1996mdl}. 
The inflaton $\phi=\sqrt{\tau_3}\varphi$ measures the separation distance between the 2 branes. The gravitational and the attractive $D3$-charge forces bring them together. Inflation happens as $\phi$ rolls down along the $T=0$ valley. The mass of $T$ is given by
\begin{equation}\label{tmass}
m_T^2(\phi)=\pi g_s\left(\frac{\phi^2}{\tau_3} -\frac{\pi} {(2 \ln 2)M_A^2}\right)V(\phi) \simeq
2 \pi g_s (\phi^2 - \phi_c^2)
\end{equation}
so $T$ becomes a tachyon when its mass square is negative, {\it i.e.}, when $\phi^2 < \phi_c^2$, where
\begin{equation}\label{Ttachyon}
\phi_c^2 = \frac{\pi \tau_3}{(2 \ln 2) M_A^2}= \frac{M_A^2}{(4 \ln 2) g_s}
\end{equation} 
 As $T$ rolls down the potential, i.e., $T \to \infty$,  fundamental $F1$-strings are produced. In addition, topological defects, i.e., the $D1$-branes, or $D1$-strings, are also produced~\cite{Jones:2002sia}.
    
 During inflation, $T=0$, so the  inflaton potential $V(\phi)=U(\phi, 0)$. At large distances (large enough $\phi$), $V(\phi)$ reduces to,
 with the $D3$ charge $\mu_3=\tau_3$,
\begin{equation} \label{Vpot1}
V(\phi)= 2 \tau_3 \left(1 - \frac{\tau_3Q}{2 \phi^4} \right), \quad \quad \phi>\phi_c 
\end{equation}
where the first term is the energy stored in the $D3$-${\bar D}3$-brane pair
and the second term comes from the gravitational plus the R-R attractive forces. The parameter $Q$ depends on the warped throat geometry. To be concrete, we follow the KKLMMT model~\cite{Kachru:2003sx} and set $Q=0.17$. A change in $Q$ simply shifts the values of $\tau_3$ and $\phi_{ini}$.
Slow-roll inflation starts at $\phi_1 >\phi_{ini}$.  The power spectrum observed by COBE, WMAP and PLANCK are generated at $\phi_{ini}$. Slow roll inflation continues as $\phi$ rolls towards $\phi_f$, where the slow-roll parameter $\eta$ reaches $\eta=-1$, i.e., when the slow-roll approximation breaks down. It further rolls for about another e-fold when it reaches $\phi_c$ and inflation ends. So we shall simply slow-roll $\phi$ all the way to $\phi_c$.\footnote{In the effective field theory context, the $\phi^{-4}$ potential stops dropping as $\phi < \phi_c$. Instead $V(\phi)$ stays finite but becomes complex \cite{Sarangi:2003sg}.} 

At $\phi_{ini}$, the slow-roll parameter $\epsilon \simeq 10^{-10}$, so the ratio $r \sim 16 \epsilon$ of the tensor to scalar perturbation is negligibly small.
The model~\cite{Kachru:2003sx} yields the scalar power spectral index $n_s$ in terms of the slow-roll parameters,
\begin{equation} 
n_s = 1 +2 \eta +6\epsilon \simeq 1-5/3N_e  
\end{equation}
Note that the value of $n_s$ is independent of the value of $Q$ or $D3$-brane tension $\tau_3$.
 Comparing it to the measured PLANCK value~\cite{Planck:2016}, 
 \begin{equation} \label{PLANCKns}
 n_s = 0.968\pm 0.006
 \end{equation}
 we find the best fit value to be $N_e\simeq 52$, well within the $N_e\simeq 50 - 60$ range expected for any viable inflationary universe model.
 
As $\phi \le \phi_c$, $T$ becomes tachyonic and starts rolling to infinity; physically, the branes start to collide and annihilate each other. The energy (i.e., $2\tau_3$) released goes to $F1$- and $D1$-strings when their massless and light modes heat up the universe. Some long $F1$-strings and $D1$-strings (with tension of order of $10^{-10} -10^{-8}$ in units of $G_N^{-1}$) form a cosmic string network~\cite{Jones:2002cv,Sarangi:2002yt,Copeland:2003bj}. Their cosmological signatures provide a crucial test of the model.

Before going to the double $D3$-${\bar D}3$-brane inflation model, one may wonder if the single $D3$-${\bar D}3$-brane inflation model can produce PBHs? In Appendix~\ref{appA}, we argue that the tunnelling process (see Fig.~\ref{figureNick}) can indeed do so, but the efficiency is low and we doubt that supermassive PBHs can be produced that way.

\section{The Double $D3$-${\bar D}3$-Brane Inflation Model}\label{double}

\subsection{The Model}

Although the double $D3$-${\bar D}3$-brane inflation model is a simple generalization of the above single $D3$-${\bar D}3$-brane inflation model, the resulting $n_s$ is sensitive to the specific setup of the model. Here we show that the idealized setup (Fig.~\ref{DoubleFig}) is consistent with the PLANCK data~\cite{Planck:2016}, while an alternative (unfavorable) setup is inconsistent with the PLANCK data. In a realistic setup, the $\phi_2$-pair is not expected to be aligned with the $\phi_1$-pair. As an example, let us move the $\phi_2$-pair away from the line; here, the $D1$-strings are produced in the presence of a dipole force, substantially weakening the $G_5$ strength.

The general setup is described in Appendix~\ref{appB}. Here, we first consider an idealized version of the double $D3$-${\bar D}3$-brane inflation model, where all the branes are lined up along a line, as shown in Fig.~\ref{DoubleFig}. Since the attractive gravitational force precisely cancels the repulsive $D3$-force between the two $D3$-branes and between the two ${\bar D}3$-branes, we are left with 
\begin{align}\label{D-model}
V(\phi)=V(\phi_1, \phi_2)=&V_0 + V_1(\phi_1) +  V_2(\phi_2) + V_3(\phi_1, \phi_2) \\\nonumber
=&4 \tau_3 - \tau_3^2Q/\phi_1^4 -  \tau_3^2Q/\phi_2^4 -  2\tau_3^2Q_3/(\left(\phi_1+\phi_2)/2\right)^4 
\end{align}
where we have set the distance between the 2 $D3$-branes equal to the distance between the 2 ${\bar D}3$-branes, i.e., namely,  $(\phi_1-\phi_2)/2>0$ and $Q_3=Q$. Here, the $\phi_2$-pair collide and annihilate before the $\phi_1$-pair. When necessary, we shall take $Q=0.17$.
  
 In an alternative setup, the 2 pairs of $D3$-${\bar D}3$-branes, i.e., the $\phi_1$-pair and the $\phi_2$-pair, are far enough apart so we can ignore the interaction between the 2 pairs, i.e., simply ignore the $V_3$ term, or, set $Q_3=0$.  As we shall see, $Q_3 \simeq 0$ is inconsistent with the PLANCK data. This is good news, as the annihilation of the $\phi_2$-pair in the presence of the $\phi_1$-pair (i.e., in the presence of a non-zero $G_5$) is crucial for the dynamics we are interested in.  

The Hubble parameter is
\begin{equation}\label{eq:H2}
H^2=V(\phi)/3M_P^2
\end{equation}
The inflations $\phi_j$ obeys 
$${\ddot \phi}_j +3H{\dot \phi}_j + \frac{dV}{d \phi_j}=0$$
In the slow-roll approximation, $\ddot \phi_j$ may be dropped, so we have
\begin{equation}\label{clock}
{\dot \phi}_j = -\frac{1}{3H} \frac{dV}{d \phi_j} = - \frac{1}{3H} (V_j'+V_3')
\end{equation}
where the prime indicates derivative with respect to $\phi_j$.
Since $V$ and so $H$ varies little, we shall take them to be constant.   The power spectrum ${\cal{P}}(k)$ at $k=aH$ (where $a$ is the cosmic scale factor) is given by
\begin{equation}
{\cal{P}}(k) \simeq \frac{1}{2 \epsilon M_P^2}\left(\frac{H}{2 \pi}\right)^2 \big|_{aH=k} =\frac{1}{24\pi^2M_P^4} \frac{4 \tau_3}{\epsilon} \big|_{aH=k}
\end{equation}
where
\begin{equation}\label{epsilon}
\epsilon=-{\dot H}/H^2=\epsilon_1 + \epsilon_2= \frac{M_P^2}{2V^2}\left[(V_1'+V_3')^2 +(V_2'+V_3')^2\right] 
\end{equation}
The scalar power spectrum at the pivot scale $k=aH=0.05 \,Mpc^{-1}$ as measured in the CMB by COBE~\cite{COBE:1992syq} and PLANCK~\cite{Planck:2016}, 
\begin{equation}\label{PLANCK}
{\cal{P}}(k) = 4.7 \times 10^{-9}
\end{equation}

In the single inflaton case, we use the rolling of the inflaton as the clock during the inflationary epoch. Here,  the $\phi_2$ brane pair will annihilate first. Since $\phi_2$ is an open string mode that measures the separation distance between the $D3$-brane and the ${\bar D}3$-brane, $\phi_2$ as a physical degree of freedom disappears after the $\phi_2$-brane pair annihilation. 
So we should use $\phi_1$ as the clock, as it continues to roll until the $\phi_1$-pair collides. Now, the annihilation of the $\phi_2$-pair releasing its $2\tau_3$ energy to heat up the universe, which evolves a while until the remaining $2\tau_3$ energy dominates over the radiation and matter released by the $\phi_2$-brane pair annihilation. Ignoring this transition period, we introduce an effective $\phi_1=\phi_{1t}$ as the transition point, when one inflationary phase transits to the second inflationary phase. Using eq.(\ref{clock}) for $\phi_1$ and noting that $\delta N \simeq H \delta t$, we have
\begin{equation}\label{NeN2}
N_e =  (N_e-N_2) + N_2 \simeq \frac{1}{M_P^2}\left[\int^{\phi_{1t}}_{\phi_c}  \frac{2 \tau_3}{V_1'} d\phi_1
 +\int^{\phi_{1, ini}}_{\phi_{1t}}  \frac{4 \tau_3}{V_1' +V_3'}d\phi_1\right]
\end{equation}
where $N_e$ is the number of e-folds covered by the $\phi_1$-pair and $N_2$ is the number of e-folds covered by the $\phi_2$-pair.
Here, $\phi_{1, ini} >\phi_{1t}$ and $N_e > N_2$; that is $\Delta N=N_e-N_2$ is the remaining e-folds the
$\phi_1$-pair goes through after the annihilation of the $\phi_2$-pair.

The power spectrum index $n_s$,
$$n_s -1 =\frac{d \,{ \ln \cal{P}}(k)}{d \,{\ln k}}$$
To compare to the observations, we need to find $d\epsilon/d{\ln k}$. 
Since $H$ varies little, we have
$$d{\ln k} =d(aH)/aH =da = H dt= - 3H^2/(V_1' +V_3') d\phi_1$$
so
\begin{equation}\label{eta-3}
\frac{d\epsilon}{d\ln k} =-M_P^2\frac{(V_1' +V_3')}{V}\frac{d\epsilon}{d\phi_1}
\end{equation}
where we shall simply set $V=4 \tau_3$.
Applying the model (\ref{D-model}) to eq.(\ref{epsilon},~\ref{eta-3}),
we obtain
\begin{align}
\epsilon_1=&\frac{M_P^2\tau_3^2}{2} \left[Q/\phi_1^5 + Q_3/((\phi_1+\phi_2)/2)^5\right]^2 \\\nonumber 
\epsilon_2=&\frac{M_P^2\tau_3^2}{2} \left[Q/\phi_2^5 + Q_3/((\phi_1+\phi_2)/2)^5\right]^2  \\\nonumber 
\eta_1=&-5M_P^2\tau_3Q/\phi_1^6 + \eta_3 \\\nonumber 
\eta_3=&-\frac{5M_P^2\tau_3}{2} Q_3/((\phi_1+\phi_2)/2)^6 \\
\eta=&\frac{2}{\epsilon}\left[\epsilon_1 \eta_1 + \sqrt{\epsilon_1\epsilon_2} \eta_3\right] \\\nonumber 
\end{align}
The model has essentially 3 parameters: \\
$\bullet$ the $D3$-brane tension $\tau_3$; \\
$\bullet$ the number of e-folds $N_e$ for  the initial $\phi_{1, ini}$ to reach $\phi_c$ when the $\phi_1$-pair inflation ends; \\
$\bullet$  the number of e-folds $N_2$ for $\phi_2$ to evolve from $\phi_{2, ini} \to \phi_c$. Here $N_e > N_2$.

\begin{figure}
 \begin{center}
  \includegraphics[width=5in]{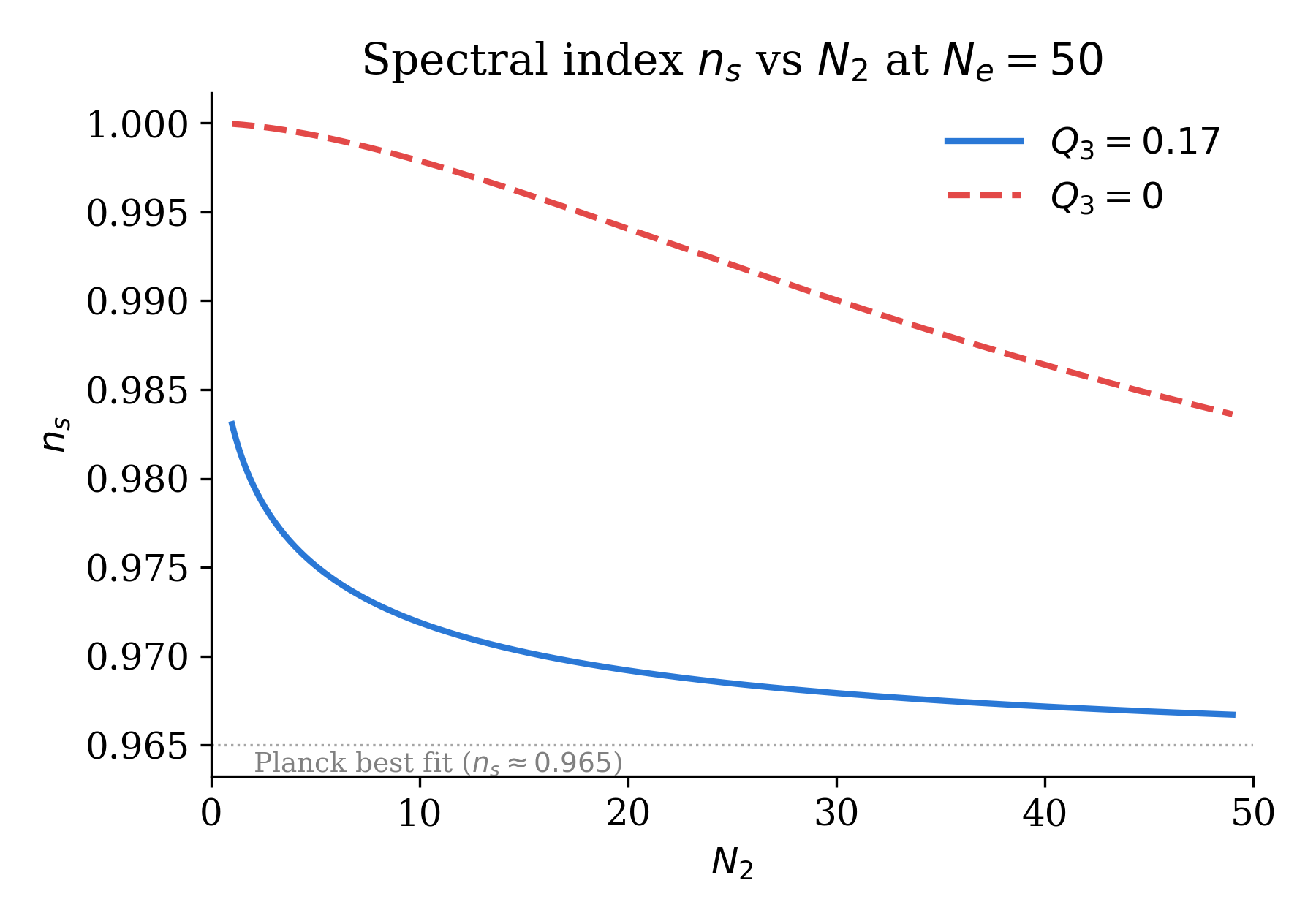}
  \caption{The spectral power index $n_s$ as a function of the number of e-folds $N_2$ of the $\phi_2$-pair in the double $D3$-${\bar{D}3}$-brane inflation model. Here, the number of e-folds of the $\phi_1$-pair is set at $N_e=50$. The blue curve is for the case when the 2 $D3$-${\bar{D}3}$-brane pairs are aligned along a line, as shown in Fig.~\ref{DoubleFig}, where $Q_3=Q$. The red dash curve is for the case when the 2 brane pairs are separated far enough apart so that $Q_3\simeq 0$. The $Q_3=0$ case is inconsistent with the PLANCK data~\cite{Planck:2016}.} 
  \label{Blue}
  \end{center}
    \end{figure}

For $\phi_{2, ini}$ close to $\phi_{1, ini}$, that is $N_2$ close to $N_e$, the 2 integrands in eq.(\ref{NeN2}) are comparable so we shall simply ignore this difference. Converting $\phi_{j,ini}$ to e-folds, we have, in a reasonable approximation,
\begin{align}\label{initialphi}
\phi_{1, ini}^6=&12 M_P^2\tau_3 QN_e \\\nonumber 
\phi^6_{2, ini}=&12M_P^2\tau_3 QN_2 \\\nonumber
\phi_{1t}^6=&12M_P^2\tau_3 Q(N_e-N_2)
\end{align}
It is convenient to introduce
\begin{align}
r=\epsilon_1/\epsilon_2=&\left[\frac{N_e^{-5/6} + \nu^{-5}}{N_2^{-5/6} + \nu^{-5}}\right]^2 \\\nonumber 
\nu=&\frac{N_e^{1/6}+N_2^{-1/6}}{2} \\\nonumber 
n_s-1=& 2 \eta = \frac{4}{r+1}\left[r\eta_1 + \sqrt{r}\eta_3\right]
\end{align}
The resulting power spectrum index $n_s$ is shown in Table~\ref{tab:ns_tau_eps} and Fig.~\ref{Blue} for $Q_3=Q$, $N_e=50$ and $N_e>N_2\ge 1$. 

\begin{table}[htbp]
\centering
\begin{tabular}{cccc}
\hline\hline
$N_2$ & $\tau_3/M_P^4$ & $\epsilon$ & $n_s$ \\
\hline
 1  & $6.81 \times 10^{-14}$ & $2.45 \times 10^{-7}$ & 0.9831 \\
 5  & $2.28 \times 10^{-15}$ & $8.20 \times 10^{-9}$ & 0.9751 \\
10 & $6.29 \times 10^{-16}$ & $2.26 \times 10^{-9} $ & 0.9719 \\
20 & $1.99 \times 10^{-16}$ & $7.14 \times 10^{-10}$ & 0.9692 \\
30 & $1.09 \times 10^{-16}$ & $3.91 \times 10^{-10}$ & 0.9679 \\
40 & $7.32 \times 10^{-17}$ & $2.63 \times 10^{-10}$ & 0.9672 \\
49 & $5.63 \times 10^{-17}$ & $2.02 \times 10^{-10}$ & 0.9667 \\
\hline\hline
\end{tabular}
\caption{$\tau_3/M_P^4$, $\epsilon$ and $n_s$ as functions of $N_2$, at fixed $N_e=50$, obtained by solving $\tau_3/M_P^4 = 6\pi^2 P(k)\,\epsilon(\tau_3/M_P^4)$
self-consistently with $Q_3=Q=0.17$ for the double $D3$-${\bar{D}3}$-brane inflation model~(\ref{D-model}). The $n_s= 0.968\pm 0.006$ value from PLANCK~\cite{Planck:2016} constrains $N_2$ to the range $49 \ge N_2 \ge 7$, with $N_2 \simeq 30$ for the best fit.}
\label{tab:ns_tau_eps}
\end{table}

The PLANCK value for the power spectral index is $n_s = 0.968 \pm 0.006$~\cite{Planck:2016}. This yields a weak constraint on the value of $N_2$ : $N_e > N_2 \ge 7$, where the best fit value $N_2 \simeq 30$ is in excellent agreement with the PLANCK value.
We have chosen $N_e=50$ as a benchmark. Shifting $N_e$ up by one unit shifts $n_s$ up by $0.0007$ only.

Following eq.(\ref{eq:tension}) we obtain the $D1$-string tension $\tau_1$ and the $F1$-string tension $\tau_F$  to be
$$G_N \tau_1 \simeq \frac{1}{\sqrt{g_s}} \times 10^{-9},  \quad \quad \tau_{F} = g_s \tau_1 $$
Note that $D1$-strings and $F1$-strings can form bound states.

Notice that, if the two brane pairs are far enough separated, $Q_3 \simeq 0$, so
$$n_s(Q_3=0) = 1 - \frac{5}{3N_e} \cdot \frac{1}{1+ (N_e/N_2)^{5/3}}$$
This yields $n_s$ inconsistent with observation, as shown as the red dash curve in Fig.~\ref{Blue}.  As pointed out earlier, it is crucial for our model that the $D3$ force is present in the region where the first brane pair annihilates. Only a vanishingly small dipole $D3$ force is present when the two brane pairs are far apart. Fortunately, this unfavorable setup is ruled out by the PLANCK data~\cite{Planck:2016}.

\subsection{After the annihilation of the first $D3$-${\bar{D}3}$-brane Pair}

In the single $D3$-${\bar{D}3}$-brane inflation model, the $D3$-brane annihilates with the ${\bar{D}3}$-brane, releasing the brane energy $2 \tau_3$ to the strings (both $D1$ and $F1$-strings), where some of the light modes of the strings behave as fundamental particles. As a result, the universe enters the hot big bang phase. 

In the double $D3$-${\bar{D}3}$-brane inflation model, there is a big difference when 
the first inflationary phase ends at $\phi _2=\phi_c$ and $T_2 \to \infty$ as the $D3$-brane annihilates with the ${\bar{D}3}$-brane, releasing the brane energy density $2 \tau_3$. Here, the annihilation happens in the presence of the remaining $D3$-${\bar{D}3}$-brane pair, which generates a background $D3$ charge field strength. As we shall now discuss, the presence of this 5-form field strength substantially alters the underlying physics. 
In comparison to the single $D3$-${\bar{D}3}$-brane inflation model, it is convenient to introduce the mass scale $M_A$ here
\begin{equation}
M_A \simeq (2 \pi g_s)^{-1/4} \,\cdot\, 10^{-4}M_P
\end{equation}
This is comparable to the corresponding $M_A$ in the single $D3$-${\bar{D}3}$-brane inflation model.

\subsection{The 5-form field strength $G_5$}

Just like a charged particle (e.g., an electron) is charged under a $1$-form potential $A_1$ (e.g., the electromagnetic gauge field $A_{\mu}$) and generates a $2$-form field strength $F_2=dA_1$ (e.g., the electric field), a $Dp$-brane is charged under a $(p+1)$-form potential $C_{p+1}$ with a $(p+2)$-form field strength $G_{p+2}=dC_{p+1}$. For $p=3$, the $D3$-attractive force follows from a $D3$-brane charged under a $4$-form potential $C_4$ with a $5$-form field strength $G_5$. This leads us to consider the 5-form field strength $G_5$ associated with the $\phi_1$ brane pair. In short, we expect $|G_5|/M_A$ to be very small but not vanishingly small.

Following eq.(\ref{Vpot1}), the attractive Ramond-Ramond (R-R) potential between the $D3$-brane and the ${\bar D}3$-brane for the $\phi_2$-pair is given by
\begin{equation}
V_{RR}(\phi) =\tau_3C_{txyz}= -\frac{\tau_3^2Q}{\phi_1^4} 
\end{equation}
where the $3$-branes span the $xyz$ directions.
Let the $D3$-${\bar D}3$-brane pair lie along the $\hat w$ direction inside the 6 extra dimensions.
The 5-form field strength $G_5$ along the $\hat w$ direction is
 \begin{equation}\label{G5}
G_{twxyz}(\phi)= dC = \frac{4\tau_3 Q}{\phi_1^5} M_A^2 \hat w = f \hat w
\end{equation}

The magnitude of the $G_5$ field strength is biggest at $\phi_c$ (\ref{Ttachyon}), 
 \begin{equation}\label{fmax}
 f_{max} /M_A = (\phi_c) \simeq 0.1 g_s^{3/2} \sim 10^{-2}
 \end{equation}
just before the collision and annihilation of the $\phi_1$-pair. When $\phi_2$-pair collides, $\phi_1=\phi_t$, so $f/M_A \simeq 10^{-3}$. So $f$ increases to $f_{max}$ and then vanishes after the collision/annihilation of the $\phi_1$-pair.
However, $f_{max}$ in this idealized case is probably unrealistically big.

Even in the idealized case where all the branes are aligned, there are 3 distinct possibilities. Let us consider: \\
$\bullet$ $D3 - (D3-{\bar D}3) - {\bar D}3$\\
$\bullet$ $(D3-{\bar D}3) - D3 - {\bar D}3$ \\
$\bullet$ $(D3-{\bar D}3) - {\bar D}3 - D3$ \\
where the bracketed $(D3-{\bar D}3)$ pair annihilates first.
Here, the first case (shown in Fig.~\ref{DoubleFig}) is the one already discussed. In this case, shifting the $\phi_2$-pair away from the symmetric point splits the 2 terms in $V_3(\phi)$~(\ref{D-model}), but has little effect on the resulting $n_s$ value.
Of the 4 interacting terms in $V(\phi)$~(\ref{D-model}), only the $V_3(\phi)$ term is different in the other 2 cases. So, again, the $n_s$ constraint from PLANCK is not very restrictive, unless the 2 brane pairs are so far apart that 2 of the 4 terms drop out (i.e., $Q_3=0$).  On the other hand, the $D1$-strings produced are in the presence of the $D3$ dipole force, which is substantially weaker than the $f_{max}$ in the first case. Let us get an order-of-magnitude estimate on the range of values of $f$ one may encounter in the model. Let the distance of the $\phi_2$-pair from the $\phi_1$-pair $\kappa$ be. Not dropping the $V_3(\phi)$ term in $V(\phi)$~(\ref{D-model}), we keep $\kappa \sim \phi_{1,ini}$. As $\phi_1 \to \phi_c$, the $G_5$ field strength felt by the $D1$-strings now comes from the dipole force (with dipole moment $\tau_3\phi_c$),  
\begin{equation}\label{typicalf}
f/M_A \sim \tau_3QM_A\phi_c/\kappa^6 \sim 10^{-10}
\end{equation}

In general, the 2 brane pairs are not aligned in a line. So the value of $f$ felt by the $D1$-strings is very sensitive to the positions and orientation of the branes. 
As an example, let us displace the $\phi_2$-pair away from the line defined by the $\phi_1$-pair by a distance of ${\Delta}$. Then the $V_3(\phi)$ (i.e., the $Q_3$) term in $V(\phi)$~(\ref{D-model}) is replaced by
$$V_3 (\varphi) = \frac{2\tau_3^2Q}{\varphi^4}, \quad \quad \varphi^2 = (\frac{\phi_1+\phi_2}{2})^2 + \Delta^2$$  
where $\Delta=0$ reduces to the idealized case with $Q_3=Q$. On the other hand, if $\Delta  \gg (\phi_1+\phi_2)/2$, then the $V_3(\phi)$ term essentially drops out.
This reduces to the (unfavorable) $Q_3=0$ case.  
To maintain consistency with the PLANCK data, we require
$(\phi_{1, ini}+\phi_{2, ini})/2 \gtrsim \Delta$. This also leads to $f \ll f_{max}$. 

In terms of the value of $f$, the idealized case and the $Q_3=0$ case are the 2 limiting cases,
with $f_{max}$~(\ref{fmax}) and $f=0$ respectively. In general, the value of $f$ is expected to be much smaller than $f_{max}$ but non-zero. We shall use $f$~(\ref{typicalf}) as a benchmark, knowing that $f$ can be bigger or much smaller than this value. Although $f$ varies with time, we shall discuss the formation of dielectric $3$-branes in the presence of a constant $f$.

 \section{Dielectric $3$-branes}\label{diDp}

Recall a dielectric in electromagnetism, where an external electric field induces a separation of charges in a neutral material, generating an electric dipole. An analogous situation happens here. As pointed out by Myers~\cite{Myers:1999ps} (and an extension~\cite{Johnson:2003cvf}), the 5-form field strength leads to the formation of a dielectric $D3$-brane (Diel-3-brane) in the presence of ${\cal{N}}$ D1-strings, due to the non-commutative geometric properties of the $D1$-strings. Such a Diel-3-brane is neutral in $D3$-brane charge but has a dipole moment. Since the production of Diel-3-branes is a crucial feature of this model, we shall review the basic idea behind their formation.

 Let us discuss 2 possible approaches to view the formation of a Diel-3-brane in the double $D3$-${\bar D}3$-brane inflation model : \\
   $\bullet$ Dual of the ($p=0 \to p+2=2$) case. We start in Type IIA string theory where a collection of ${\cal{N}}$ $D0$-branes forms a dielectric $D2$-brane with $S^2$ (2-sphere) topology. Under a T-duality transformation ($p=0 \to p=1$), we end up with a Diel-3-brane with ${\cal{R}} \times S^2$ topology in Type IIB string theory. \\
   $\bullet$ The ($p=1 \to p+2=3$) case. We start directly in Type IIB string theory, where the Diel-3-brane has ${\cal{R}} \times S^2$ topology. 
   
   Here we shall first review the general $p \to p+2$ case in flat spacetime and then apply it to the $p=0$ and the $p=1$ cases. Although Myers discusses only the $p=0$ case \cite{Myers:1999ps}, the generalization from the $D0$-brane case to the general $p$ case is quite straightforward~\cite{Johnson:2003cvf}. Here, our final interest is the $p=1$ case.

\subsection{Review of Dielectric $D(p+2)$-branes}

For general $p$, consider only the non-trivial components of $G_{(p+4)}$,  
$$G_{t1...pijk}=G_{tijk}=-2f \epsilon_{ijk} \quad i,j,k  \in \{1,2,3\}$$
where $\epsilon_{ijk}$ is the antisymmetric Levi-Civita symbol.
Here a $Dp$-brane is point-like in the $ijk$ directions (i.e., $x^i \, (i=1,2,3)$) and extends in the $p$ other directions (for our case, the $p=1$ direction will be specified below). Since there is no structure there, the indices $1...p$ in $G_{(p+4)}$ are suppressed.
The ${\cal{N}}$ $Dp$-branes are described by ${\cal{N}} \times {\cal{N}}$ matrix $\Phi^i$, $(i=1,2,3)$. Here,  the diagonal elements $x^i_n$ $(n=1,2,..,{\cal{N}})$ in $\Phi^i/M_A^2$ are the co-ordinates of the branes in the $ijk$ directions. For a static configuration in flat space-time, the scalar potential takes the form
\begin{equation}\label{myers}
V(\Phi)= {{\cal{N}}} \tau_p - \frac{\tau_p}{4 M_A^4} {\rm Tr}([\Phi^i, \Phi^j]^2) - \frac{\mu_p}{3 M_A^4}{\rm Tr}\left(\Phi^i \Phi^j \Phi^k \right) G_{tijk}
\end{equation} 
where the coupling $\mu_p=\tau_p$. The second term and its non-abelian feature can be viewed as inheriting from the reduction of 10-dim. Yang-Mills theory.
This potential yields
$$\frac{\partial V(\Phi)}{\partial \Phi^i}\propto [[\Phi^i, \Phi^j], \Phi^j] + f \epsilon_{ijk}[\Phi^j, \Phi^k]=0$$
When $f=0$, a simple (supersymmetric) solution is $[\Phi^i, \Phi^j]=0$, where $\Phi^i= {\rm diag} (x^i_1, x^i_2, ... , x^i_{\cal{N}})$, i.e., the $Dp$-branes are parallel and $V(\Phi)={\cal{N}}\tau_p$. 

However, this solution is unstable as $f$ is turned on.  A lower energy solution, namely, the ground state, is given by the ${\cal{N}} \times {\cal{N}}$ irreducible representation of $SU(2)$,
$$\Phi^i=-if\Pi^i, \quad \quad [\Pi^i, \Pi^j]=2 i \epsilon_{ijk} \Pi^k$$
and 
\begin{equation}
V(\Phi)={\cal{N}} \tau_p \left[1 - \frac{f^4({\cal{N}}^2-1)}{12M_A^4} \right]
\end{equation}
In this case, the solution has the topology of ${\cal{R}}^p \times S^2$, where ${\cal{R}}^p$ part is where the ${\cal{N}}$ $Dp$-branes are extended. This solution corresponds to the non-commutative fuzzy two-sphere~\cite{Madore:1991bw,Chu:2001xi}.
The radius of the spherical $S^2$ is, for ${\cal{N}} \gg 1$,
\begin{equation} \label{RN}
R_d = \frac{f}{2M_A^2}\sqrt{({\cal{N}}^2-1)}\simeq  \frac{{\cal{N}}}{2}\frac{f}{M_A^2}
\end{equation}
Note that the binding energy is proportional to $f^4{\cal{N}}^2$, reflecting its sensitivity to $f$ and ${\cal{N}}$.
%For any given ${\cal{N}}$, the irreducible representation of $SU(2)$ yields lowest energy state. 
 
 \subsection{Dielectric $D3$-brane via T-duality}

Recall that Type IIA string theory has only even $Dp$-branes while Type IIB string theory has only odd $Dp$-branes, while they are related by a T-duality transformation. 
 Let us apply the above discussions to the case where $p=0$ in Type IIA string theory. Here, we have 
 a $D$2-brane moving towards a parallel ${\bar D}$2-brane.  The $D$2-${\bar D}$2-brane pair, lying in the $xy$-plane, move towards each other in the $w$-direction, one of the extra dimensions. Their collision yields a sea of $D0$-branes. A collection of ${\cal{N}}$ $D0$-branes form a $S^2$ sphere, namely a Diel-2-branes. This $S^2$ sphere lies in the $xyz$ directions. Now, we perform a T-duality transformation in the $z$-direction, so \\
 $D0$-brane $\to$  $D1$-string lying in the $z$-direction \\
 $D2$-brane (extending in the $xy$-directions) $\to$  $D3$-brane (in the $xyz$-directions)\\
 The R-R potential $C_{txy}$ $\to$ $C_{txyz}$ \\
 The R-R field strength $G_{txyw}$ $\to$ $G_{txyzw}$\\
 the $D$2-${\bar D}$2-brane pair become a $D3$-${\bar D}3$-brane pair;
 and the $D0$-branes become $D1$-strings lying in the $z-$directions. 
 A $D3$-brane is charged under a R-R 4-form potential $C_4$.  In the presence of the $D3$-${\bar D}3$-brane pair, there is a background  R-R 5-form field strength $G_5=dC_4$. So a dielectric $2$-brane with the shape of a $S^2$ sphere becomes a dielectric $3$-brane with ${\cal{R}} \times S^2$ topology, where ${\cal{R}}$ extends in the $z$-direction.

 \subsection{Dielectric $3$-brane in Type IIB theory}

Let us now go back to the original $p=1$ case. 
As inferred from matrix model analysis, a Diel-3-brane is a neutral 3-brane with $D1$-strings bound to it~\cite{Kabat:1997im,Rey}.
The above analysis shows how a collection of $D1$-strings can form a Diel-3-brane in the presence of the $G_5$ field strength $f$. However, in the brane inflation model, the $D1$-strings produced in the $D3$-${\bar D}3$-brane pair annihilation may be too energetic to settle down to form Diel-3-branes. In our scenario, it may be more efficient to form Diel-3-branes directly when $D3$-${\bar D}3$-brane pair collides; that is, a Diel-3-brane can be formed without first starting with only a sea of $D1$-strings.

In our model, the first $D3$-${\bar D}3$-brane pair annihilation can directly lead to the formation of Diel-3-branes when fragments of the $D3$-brane collide and merge with the ${\bar D}3$-brane (or vice versa).
Here, the underlying dynamics of a Diel-3-brane can also be viewed from the $3$-brane perspective. 
Reverting to the general $p$ case, we have a $(p+2)$-brane bounded with ${\cal{N}}$ $Dp$-branes, which takes the form of ${\cal{R}}^p \times S^2$. The potential takes the form~\cite{Myers:1999ps,Johnson:2003cvf,Emparan:1997rt},
\begin{equation}\label{p+2}
V(R) = 4 \pi \tau_{(p+2)} \left([R^4+ \frac{{\cal{N}}^2}{4M_A^4}]^{1/2} -\frac{2f}{3} R^3 \right) \\
\end{equation}
Remembering that $ 2\pi \tau_{(p+2)}=M_A^2 \tau_p$, we have the solution when $f=0$:
$V(R=0)\simeq {\cal{N}}\tau_p$ at $R=0$, corresponding to flat ${\cal{N}}$ $Dp$-branes.  
However, when $f$ is turned on, this solution is unstable.  Setting $dV(R)/dR=0$ now yields,
\begin{equation}\label{Rpm}
R_{\pm}^2 = \left(1 \pm \sqrt{1 - {\cal{N}}^2f^4/M_A^4}\right)/2f^2
\end{equation}
where $R_{-}$ minimizes $V(R)$. In the large ${\cal{N}}$, small $f$ regime, it reduces to the same $R_d$ as given in eq.(\ref{RN}). The $R_{+} \simeq f^{-1}$ indicates a local maximum, i.e., the barrier separating $R_{-}$ from runaway growth. Physically, as the $D3$-brane collides with the ${\bar D}3$-brane, there is no limit to the size of a fragment of the $D3$-brane merging with a fragment of the ${\bar D}3$-brane. It is the sea of $D1$-strings in the presence of $f$ that stabilizes some of the merged fragments into Diel-3-branes. 

Applying the constraint for $R_-$~(\ref{Rpm}) on the radius $R_d$~(\ref{RN}), 
$$ 1 >  {\cal{N}} \left(\frac{f}{M_A}\right)^2  \quad \to \quad  \frac{1}{2f} > R_d$$
For typical values of $f$~(\ref{typicalf}), one also finds ${\cal{N}} < 10^{20}$.
Note that the value of $f$ varies as the $\phi_1$-pair continues to inflate after the $\phi_2$-pair collision.
Depending on the position, the orientation and the value of $\phi_{2, ini}$ of the $\phi_2$-pair relative to the $\phi_1$-pair, subject to the constraint imposed by the $n_s$ value from PLANCK~\cite{Planck:2016}, the bound on ${\cal{N}}$ can be further relaxed.

\section{Properties of Diel-3-branes}\label{diD3M}

 Now we are ready to consider what and how the Diel-3-branes can be realized in the model. 

\subsection{A 3-Ball Diel-3-brane}

 Let us consider the Diel-3-brane obtained via T-duality, where ${\cal{R}}$ is along the $z$-direction.
Energetically, instead of stretching outside the $S^2$ sphere with no endings, the $D1$-strings stretch inside the $S^2$ sphere, with its two ends ending on the $S^2$. This results in a solid 3-ball. 
In the case where ${\cal{R}}$ extends in one of the extra dimensions, the $D1$-strings can simply shrink to a minimum length, where its two ends live on the $S^2$ sphere. This also leads to a solid 3-ball. Order-of-magnitude wise, such a 3-ball with radius $r_3$ has mass ${\cal{M}}_3 \sim (4\pi r_3^3)/(3) \tau_3$.  
For ${\cal{N}} \sim 0$ and $f \sim 0$, eq.(\ref{p+2}) reduces to
\begin{equation}\label{RbigN}
V(R)\simeq 4 \pi R^2 \tau_3
\end{equation}
yielding a 3-ball mass
\begin{equation}\label{3mass}
{\cal{M}}_3 \simeq \int_0^{r_3} V(R) dR= \frac{4\pi r_3^3}{3} \tau_3
\end{equation}
as expected.  Let us make an order-of-magnitude estimate of $r_3$.
To minimize ${\cal{M}}_3$, we use the approximate $V(R)$ (\ref{p+2}) for ${\cal{N}}/2M_A^2 < R^2$ and small $f$ to obtain
$$r_3\simeq  \frac{2 f {\cal{N}}}{3M_A^2}$$ 
Substituting this radius $r_3$ into eq.(\ref{3mass}) gives the approximate $3$-ball mass 
${\cal{M}}_3$~(\ref{3mass})  when we ignore the correction terms in eq.(\ref{p+2}). 

Next, let us compare this to the Schwarzschild radius. To satisfy the Schwarzschild radius ($r_S$) criterion, where
$$r_S= {\cal{M}}_3/4 \pi M_P^2$$
For the 3-ball to collapse to a black hole, we need
$$r_S > r_3 \quad \rightarrow  \quad r_3^2 > \frac{3}{M_A^2} (M_P/M_A)^2 $$ 
or, with $M_P/M_A \simeq  10^4$,
$$r_3 >  10^4/M_A  \quad \to \quad {\cal{M}} >  10^{9}M_P $$
This condition is easy to satisfy.

\subsection{A Dumbbell Diel-3-brane}

Since all the extra dimensions are compactified, the $D1$-strings extending in one of the extra dimensions may simply wrap around that direction. If it wraps around a cycle with length $l$, then its mass is of order ${\cal M} \simeq {\cal{N}}l\tau_1$.
As $f \to 0$, it reduces to a BPS state. Its effective mass may be substantially larger than expected if the cycle leaves the bottom of the warped throat. 

 More interestingly, consider two $S^2$s:  the $S_1^2$ sphere has ${\cal{N}}_1$ $D1$-strings stretched inside it like a 3-ball, while the other $S_2^2$ sphere has ${\cal{N}}_2$ $D1$-strings stretched inside it like another 3-ball. They are separated by a distance of $\lambda$ and there are ${\cal{N}}_{12}$ $D1$-strings stretched between them. 
 This configuration looks like a dumbbell. Since $D1$-strings and $F1$-strings can form bound states~\cite{Schwarz:1995dk,Polchinski:1995mt,Witten:1995im}, we expect the presence of $F1$-strings as well.
 
As a simple example, consider the dumbbell case with ${\cal{N}}_1={\cal{N}}_2=0$ and ${\cal{N}}_{12}={\cal{N}}$, that is, ${\cal{N}}$ $D1$-strings stretching between two same size (with radius $R_d$) spherical $S^2$ separated by a distance $\lambda$. Such a dumbbell has mass and size
$${\cal{M}}_{db}\simeq {\cal{N}} \lambda \tau_1, \quad \quad  V_{db}=\pi R_d^2 \lambda$$
so the Schwarzschild radius is
$$r_S={\cal{N}}\lambda \tau_1/4 \pi M_P^2$$
For the formation of a black hole, we need
$$ r_S > \lambda, \quad \quad r_S > R_d$$
or
$${\cal{N}} > 10^9, \quad \quad \lambda >\frac{10^9}{M_A} \frac{f}{M_A}$$
These conditions are easy to satisfy, especially when the 2 $S^2$s are far apart and $f \to 0$.

On the other hand, if ${\cal{N}}_1$ and ${\cal{N}}_2$ are large but ${\cal{N}}_{12}$ is small, one can end up with the 2 $S^2$ spheres collapsing to 2 black holes linked by stretched ${\cal{N}}_{12}$ $D1$-strings; these stretched strings tend to pull them towards each other, thus accelerating the merging of the 2 black holes. 

\subsection{A Network and Other Possibilities}

\begin{figure}
 \begin{center}
  \includegraphics[width=4.5in]{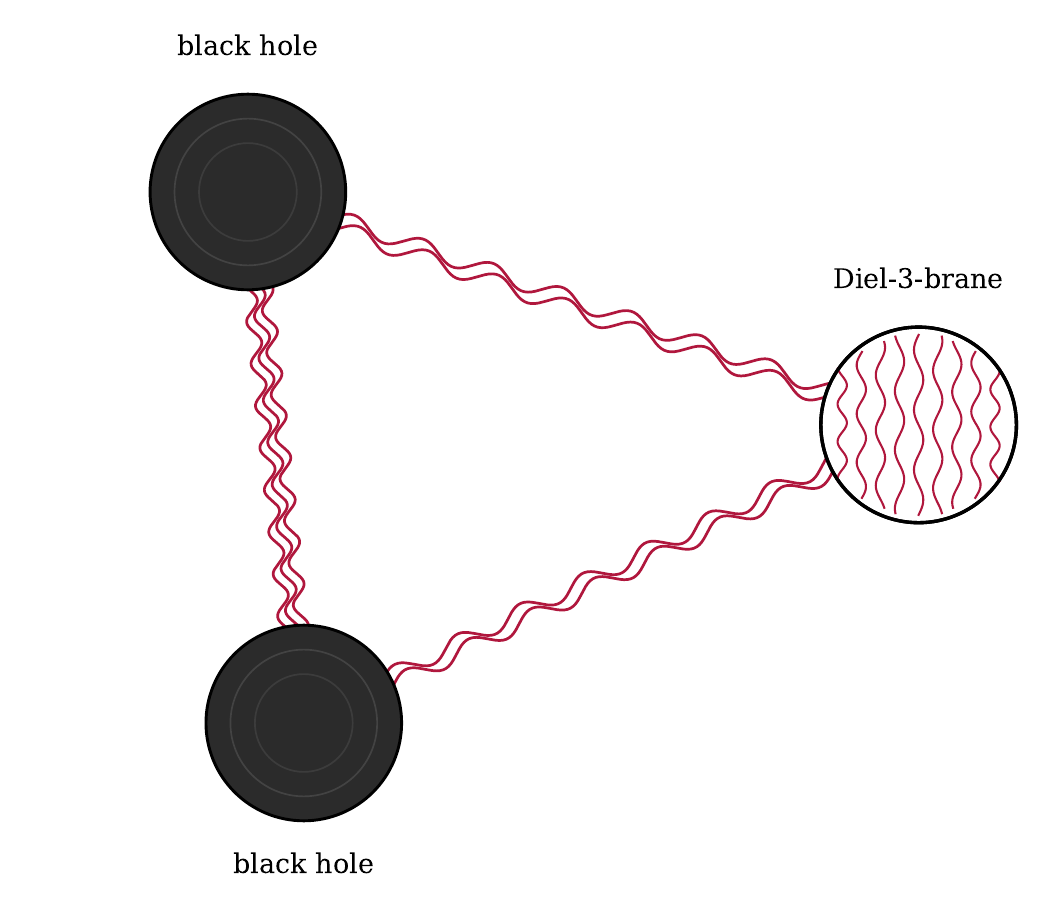}
  \caption{A simple network of $M=3$ Diel-3-branes, where two of them have already collapsed to black holes (each indicated by a grey circle). The remaining Diel-3-brane is a 3-ball with $D1$-strings stretched inside its 2-sphere. The wiggly red lines are $D1$-strings.} 
  \label{network}
  \end{center}
    \end{figure}

The above dumbbell configuration easily generalizes to a network of multiple $S^2$-spheres linked by $D1$-strings. Consider a network with $M$ $S^2$-spheres: $S^2_m$, $m=1,2,..,M$. each $S^2_m$ has ${\cal{N}}_m$ $D1$-strings stretched inside it and ${\cal{N}}_{nm}$ $D1$-strings linking it to the $S^2_n$-sphere. Her $M$ can be large.
The $D1$-strings stretching between any two $S^2$s tend to pull them towards each other. PBHs may be formed during the second inflationary epoch or after, while the merging of them due to the $D1$-string pull can happen during the radiation/matter dominating phases in the early universe. A simple $M=3$ case is illustrated in Fig.~\ref{network}. Such networks may be ubiquitous in early universe. 
 
We also expect $F1$-strings to be present inside any Diel-3-brane. They may also stretch between different Diel-3-branes. 

Spinning, and/or oscillating Diel-3-branes are also expected. Together, they provide a rich landscape.

\section{Primordial Black Hole Formation from Diel-3-branes}\label{PBHS}

Let us make a crude estimate on the feasibility of the PBH production mechanism in the model. To be precise, let us estimate the fraction of energy (out of the $2\tau_3$) that goes into Diel-3-branes.
Let $F(M)$ be the mass distribution of the Diel-3-branes formed. The fraction of total Diel-3-brane mass to the energy released $2 \tau_3$,
$$F = \int^{M_{max}}_{M_{min}} F(M) \frac{dM}{M}, \quad \quad F < 1$$
We want to check whether and under what condition this $F < 1$ constraint is satisfied. Here, there is no known theoretical upper bound $M_{max}$ on the Diel-3-brane mass as ${\cal{N}}$ is in practice unbounded. Let us take the cosmic scale factor at the end of inflation to be $a_{inf} \simeq 10^{-26}$. \footnote{If reheat is efficient, it would be $10^{-27}$ instead. Here a slightly more conservative value is chosen.}  
Assuming that Diel-3-branes are formed after the $\phi_2$-brane pair annihilates, but before the $\phi_1$-pair annihilation, $F(M)$ may be diluted by the second phase of inflation due to the $\phi_1$-brane pair, i.e, a factor of $e^{-3\Delta N}$, where $N_e-N_2 \ge  \Delta N \ge 0$. Here, the upper bound applies if the Diel-3-branes are formed immediately after the $\phi_2$-pair annihilation and the lower bound applies if the Diel-3-branes are formed and continue to grow during the second phase of inflation due to the $\phi_1$-pair.

The abundance and distribution of PBH masses can be characterized by the PBH mass function $\psi(M)$, so the total PBH mass to the dark matter fraction is
$$f_{PBH} = \frac{\Omega_{PBH}}{\Omega_{DM}} = \int^{M_{max}}_{M_{min}} \psi(M) \frac{dM}{M}$$ 
where $\Omega_{DM}\simeq 0.27$ is the measured dark matter density fraction.
We would like to find the relation between $F(M)$ and $\psi (M)$. Loosely speaking, we assume $10^{-5} > f_{PBH} > 10^{-12}$.

To be concrete, let us adopt for the dark matter the fuzzy dark matter model  ~\cite{Hu:2000ke,Schive:2014dra,Hui:2016ltb} due to an axion field~(cf.\cite{Marsh:2015xka}).  
When $H > m$, the axion field $\zeta$ is overdamped and is frozen at its initial mis-aligned value $\zeta_i$ by the Hubble friction. When $H<m$, the axion field is underdamped and its density starts to evolve like matter, with initial density $\rho_{FDM} = m^2\zeta_i^2$. For $m \simeq 10^{-22}$ eV, this happens at around $a_{FDM}\simeq 4 \times 10^{-7} \sim 10^{-6}$ with $\zeta_i \simeq 10^{17}$ GeV; \footnote{This value of $\zeta_i$ is fixed by requiring the axion, once rolling, to redshift as matter and account for all of the dark matter today: $m^2\zeta^2 a_{FDM}^3 = \Omega_{DM}\, 3H_0^2M_P^2$, i.e. $\zeta \simeq (1/m)\sqrt{3\Omega_{DM}H_0^2M_P^2/a_{FDM}^3} \simeq 10^{17}$ GeV for $\Omega_{DM}\simeq 0.27$.} 
so $\rho_{FDM}/M_P^4 \simeq 10^{-102}$. 
Here, $F(M)$ continues to be diluted as $a_{inf} \to a_{FDM}$; after which both decreases as matter density. Putting the various factors together, we have
\begin{align}
\psi (M) \simeq & A(M)\left(\frac{a_{inf}}{a_{FDM}}\right)^3 e^{-3\Delta N} \frac{2\tau_3}{\rho_{FDM}} F (M) \\\nonumber
\sim & A(M)\left(\frac{a_{inf}}{10^{-26}}\right)^3 \left(\frac{a_{FDM}}{10^{-6}}\right)^{-3} \left(\frac{10^{-1.3\Delta N}}{10^{-26}}\right)\left(\frac{\tau_3/M_P^4}{10^{-17}}\right) \left(\frac{m^2\zeta^2/M_P^4}{10^{-102}}\right)^{-1}  F(M) \\\nonumber
\sim & A(M) \frac{10^{-1.3\Delta N}}{10^{-26}} F(M)
\end{align}
where $A(M)$ is the accretion factor for PBH with mass $M$. 
Estimating the accretion factor $A(M)$ is complicated~\cite{Escriva:2022duf,Carr:2023tpt,Carr:2026hot,Zhang:2025oyl}. Depending on the mass $M$, $10^7 > A(M) \gtrsim 1$,
where we quote the upper value for $M \sim 10^3 M_{\odot}$ \cite{Yuan:2023bvh}. 

If the Diel-3-branes are formed quickly after the $\phi_2$-pair annihilation, $F(M)$ would be further diluted by $\Delta N$ e-folds.
Using the best fit value $N_2\simeq 30$ from Table~\ref{tab:ns_tau_eps}, $\Delta N =N_e-N_2=20$, we have,
\begin{equation}
\psi(M) \simeq A(M) F(M) \gtrsim F(M)
\end{equation}
Since astrophysical/cosmological constraints imply $f_{PBH} \ll 1$, one obtains a similar constraint on $F$,
$$f_{PBH} \ll 1  \quad \to \quad F \ll 1$$
On the other hand, if the Diel-3-branes take time to form and grow during the second phase of $\phi_1$-pair inflation, then $F(M)$ is not diluted by the inflation after the $\phi_2$-pair annihilation, i.e., $\Delta N \simeq 0$. However, the energy density that eventually goes to the  
Diel-3-branes behaves like radiation, so $F(M)$ is still tiny.
That is, the energy going to the production of Diel-3-branes is a very small fraction of the available energy released by the $\phi_2$-brane pair annihilation. 

As a useful trial, the extended mass function for $\psi(M)$ often used, namely, the log-normal distribution with 2 parameters $M_c$ and $\sigma$, can be applicable to $F(M)$ too,
$$F(M) \propto \frac{1}{\sqrt{2 \pi} \sigma} \exp \left[- \frac{\ln^2(M/M_c)}{2 \sigma^2}\right]$$
with a relatively large $\sigma$, where $\langle M \rangle = M_c e^{-\sigma^2/2}$. If we include the tunnelling effect described in Appendix~\ref{appA}, $F(M)$ would be a combination of 3 such log-normal distributions.

\section{Summary and Remarks}\label{Remark}

The successful $D3$-${\bar D}3$-brane inflation model is extended to 2 pairs of $D3$-${\bar D}3$-branes. In the double $D3$-${\bar D}3$-brane inflation setup, one $D3$-${\bar D}3$-brane pair's collision/annihilation copiously produces $D1$-strings in the presence of the other pair. Before its annihilation, the presence of the $D3$-charge flux lines of the second pair induces the formation of Diel-3-branes, neutral 3-branes bound with $D1$-strings. With the number ${\cal{N}}$ of $D1$-strings (or $R_d$) in eq.(\ref{RN}) relatively free to take different values, the masses of the Diel-3-branes span a wide range, although we expect only a tiny fraction of the energy released to go to Diel-3-branes. Behaving like dark matter with little or no pressure, they act as seeds of PBHs. In short, a better understanding of the extended mass function $F(M)$ for the Diel-3-branes is crucial in its confrontation with astrophysical/cosmological observations.

 Here we conclude with a few remarks.
 
 $\bullet$ Although the formation of Diel-3-branes in this inflationary universe model is very novel, it is based on well - established string theory properties and automatically follows from the double $D3$-${\bar D}3$-brane inflation scenario.

$\bullet$ Although the model is well-defined and the production of Diel-3-branes follows automatically, its dynamics is non-trivial and remains to be fully explored. The position, the orientation and the size (i.e., the value of $\phi_2$) of the $\phi_2$-pair relative to the $\phi_1$-pair determines the range of values of $f$ takes as it varies during the evolution of the model. This will  play a huge role in determining the extended mass function $F(M)$, which in turn determines the PBH mass function $\psi(M)$.

$\bullet$ A better understanding of the properties of the Diel-3-branes will be important. Since they are neutral, the Diel-3-branes may eventually decay or collapse to BPS states. However,  although Diel-3-branes are neutral, they have dipole couplings, generating a non-zero $f$ around them. Will these dipole moments survive after the $\phi_1$-pair annihilation ? How will this impact on the PBH formation ?

$\bullet$ Note that the first 2 terms in eq.(\ref{myers}) come from the Dirac-Born-Infeld action. In the above analysis, we assume the magnitude $f$ of the 5-form field strength to be constant. In the model, the 5-form field strength $f$ actually varies. So the other terms in the model may not be ignored. 

$\bullet$ It is important to note that there may be other mechanisms to produce PBHs in addition to the ones described in this paper. Even within the double $D3$-${\bar D}3$-brane inflation model, dynamics at lower energy scales (e.g., QCD dynamics) can happen.

\begin{appendices}

\section{Can PBHs be produced in the original single $D3$-${\bar D}3$-brane Inflation Model ?}\label{appA}

When inflation ends at $\phi =\phi_c$, $T \to \infty$ as the $D3$-brane annihilates with the ${\bar{D}3}$-brane, releasing the brane energy $2 \tau_3$ so  the universe enters the hot big bang phase. Before that, tunnelling through the potential barrier (as shown in Fig.~\ref{figureNick}) followed by $T \to \infty$ can also happen.
It is easy to understand the presence of this potential barrier.  For $\phi >\phi_c$, the branes are separated. A nucleation bubble forms when a bubble of the $D3$-brane reaches and annihilates in the ${\bar{D}}3$-brane (or vice versa). For this to happen, the $D3$-brane bubble has to jump (or tunnel) over the gap between the branes. This tunnelling process is heavily suppressed as the separation distance between the branes is large.
As $ \phi \to \phi_c$, the potential barrier is decreasing and so the tunnelling probability is not as suppressed.
The resulting nucleation bubble emerges before the full $D3$-${\bar{D}3}$-brane pair annihilation so it is in the presence of the background $D3$-charge force field, which can lead to the formation of Diel-3-branes.

The tunnelling probability $P(\phi) \simeq e^{-S_E}$, where $S_E$ is given in Ref.\cite{Jones:2002sia,Tye:2023rwa},
\begin{equation}\label{SE}
S_E \simeq 8 (27) \pi^5 c_1c_2^4\left(\frac{\phi -\phi_c}{M_A} \right)^2 \simeq 700\left(\frac{\delta \phi}{M_A} \right)^2
\end{equation}
where $c_1=1.5$, $c_2=0.29$ and the standard variance $\sigma =0.0267$.
 If we treat this fluctuation as a fluctuation forward in time before the end of inflation, namely, $\delta t$, then it is equivalent to a fluctuation $\delta \phi_T$. To estimate the tunnelling effect, consider the power spectrum
\begin{align}
{\cal{P}}(k) = \frac{1}{2 \epsilon M_P^2} \left(\delta \phi \right)^2, \quad 
\left(\delta \phi \right)^2 = \delta \phi_H^2 + \left(\delta \phi_T \right)^2 = \left(\frac{H}{2 \pi} \right)^2 + \left(\delta \phi_T \right)^2
\end{align}
where an order of magnitude estimate of the value for $\delta \phi_T$ is given by\footnote{or equivalently, $\int_0^\infty (\delta\phi)^2 e^{-S_E(\delta \phi)} d(\delta \phi)/\int_0^\infty e^{-S_E(\delta \phi)} d(\delta \phi)$}
\begin{equation}
\delta \phi_T^2 = \int_0^\infty \delta\phi e^{-S_E(\delta \phi)} d(\delta \phi)\simeq (0.0267 M_A)^2
\end{equation}
where the tunnelling effect is biggest around $\phi \simeq  \delta \phi_T +\phi_c$.
We find that $\delta \phi_T \gg H/2 \pi$. At $\phi_c + \delta \phi$, just before full collision,
$${\cal{P}}(k)  \sim 4 \times 10^{-5},$$
which is much bigger than its value ${\cal{P}}(k)\simeq 4.7 \times 10^{-9}$ at $N_e \simeq 50$. Due to the small size of the nucleation bubbles, the production of massive Diel-3-branes is limited. Still, the resulting PBHs produced in this way may supplement those produced in the double $D3$-${\bar D}3$-brane inflation model.

 \section{General Double $D3$-${\bar D}3$-brane Inflation Model}\label{appB}

Let us start with two $D3$-branes and two ${\bar D}3$-branes in a compactified manifold.  Even if they start with different orientations, inflation will bring them parallel to each other. So they are point-like sitting in the 6 extra-dimensions with coordinates ${\bf y}_1$, ${\bf y}_2$ for the 2 $D3$-branes and ${\bf z}_1$, ${\bf z}_2$ for the 2 ${\bar D}3$-branes.  With the $D3$ charge (coupling) equal to the $D3$-brane tension, i.e., $\mu_3= \tau_3$, the repulsive $D3$ charge force precisely cancels the attractive gravitational force between same sign branes (i.e., they are BPS w.r.t. each other); so there are only attractive potentials between the ${\bar D}3$-branes and the $D3$-branes. The effective potential can be somewhat involved (cf.\cite{Jones:2003ae}). Fortunately, during the inflationary epoch, the inflaton potential is reduced to, 
%for $W=\tau_3^2Q/M_A^8$, 
$$V=4\tau_3 -  \frac{\tau_3^2Q}{M_A^8}\left(\sum 1/ |{\bf y}_i -{\bf z}_j|^4 \right)$$
where there are only 4 terms in the sum. In the idealized case in the text, they are lined up along the $\hat w$ direction, where $({\bf y}_1, {\bf z}_1)$ for the $\phi_1$-pair and $({\bf y}_2, {\bf z}_2)$ for the $\phi_2$-pair.
The second pair will annihilate first, as $\phi_2 \propto |{\bf y}_2-{\bf z}_2| \to 0$, followed by  $\phi_1 \propto |{\bf y}_1-{\bf z}_1| \to 0$. 

If the 2 pairs are separated far apart,  
$$|{\bf y}_1-{\bf z}_2| \sim |{\bf y}_2-{\bf z}_1| \gg |{\bf y}_1-{\bf z}_1|\gtrsim |{\bf y}_2-{\bf z}_2|$$ 
this leads to weaker forces and slower evolution of the $|{\bf y}_1-{\bf z}_2|$ and the $|{\bf y}_2-{\bf z}_1|$ terms in $V$, so their contributions may be ignored in the leading approximation. In the text, we consider the limiting case when we drop the $V_3(\phi)$ term in $V(\phi)$~(\ref{D-model}) (i.e., set $Q_3=0$). In terms of the magnitude of the $G_5$ strength $f$, the idealized case and this $V_3(\phi)=0$ case are the 2 limiting cases: with $f_{max} \sim 10^{-2}$ and $f=0$ respectively. 

%If we start with more than 2 $D3$-${\bar D}3$-brane pairs, the 2 pairs model describes the physics after all except two pairs have annihilated.

\end{appendices}

\begin{acknowledgments}
		I acknowledge the use of CLAUDE AI in the preparation of this paper.
				
\end{acknowledgments}

\end{document}